\documentclass[twocolumn,showpacs,preprintnumbers,amsmath,amssymb]{revtex4}

\usepackage{graphicx}
\usepackage{dcolumn}
\usepackage{bm}
\usepackage{color}

\begin{document}

\preprint{APS/123-QED}

\title{Strong Correlation between Anomalous Quasiparticle Scattering and \\Unconventional  Superconductivity in Hidden Order Phase of URu$_2$Si$_2$\footnote{Phys. Rev. B {\bf 85}, 054516 (2012)}}

\author{Nayouki Tateiwa$^{1}$}\email[]{tateiwa.naoyuki@jaea.go.jp}
\author{Tatsuma D Matsuda$^{1}$}
\author{Yoshichika {\=O}nuki$^{1,2}$} 
\author{Yoshinori Haga$^{1,3}$}
\author{Zachary Fisk$^{1,4}$}

 \affiliation{%
$^{1}$Advanced Science Research Center, Japan Atomic Energy Agency, Tokai, Ibaraki 319-1195, Japan\\
$^{2}$Graduate School of Science, Osaka University, Toyonaka, Osaka 560-0043, Japan\\
$^{3}$JST, Transformative Research-Project on Iron Pnictides (TRIP), Tokyo 102-0075, Japan\\
$^{4}$University of California, Irvine, California 92697, USA
}
 \date{\today}

\begin{abstract}
The pressure dependent electrical resistivity of URu$_2$Si$_2$ has been studied at high pressure across the first order phase boundary of $P_x$ where the ground state switches under pressure  from  ``hidden order"  (HO) to large moment antiferromagnetic (LAFM) states. The electrical transport in URu$_2$Si$_2$ at low temperatures shows a strong sample dependence. We have measured an ultra-clean single crystal whose quality is the highest among those used in previous studies. The generalized power law ${\rho}{\,}={\,}{\rho_0}+{\,}{A_n}T{^n}$ analysis finds that the electric transport property deviates from Fermi liquid theory in the HO phase but obeys the theory well above $P_x$. The analysis using the polynomial in $T$ expression  ${\rho}{\,}={\,}{\rho_0}+{\,}{{\alpha}_1}T+{\,}{{\alpha}_2}T{^2}$ reveals the relation ${{\alpha}_1}/{{\alpha}_2}$ $\propto$ $T_{sc}$ in the HO phase. While the pressure dependence of ${{\alpha}_2}$ is very weak, ${{\alpha}_1}$ is roughly proportional to $T_{sc}$. This suggests a strong correlation between the anomalous quasiparticle scattering and the superconductivity and that both have a common origin. The present study clarifies a universality of the HO phase inherent in strongly correlated electron superconductors near quantum criticality.

\end{abstract}

\maketitle

  Tetragonal URu$_2$Si$_2$ is a heavy-fermion superconductor with a superconducting (SC) transition temperature $T_{sc}{\,}{=}$ 1.5 K at ambient pressure~\cite{palstra}.The compound undergoes a second order phase transition at $T_0{\,}{=}$ 17.5 K, and this ordered state coexists with the unconventional superconductivity of chiral $d$-wave symmetry~\cite{kasahara}.  Although many theoretical models have been proposed for the ordered state ~\cite{mydosh}, the nature of the state is still not understood and is known as ``hidden order" (HO). Applying pressure to URu$_2$Si$_2$ induces a first order phase transition from the HO to a large moment antiferromagnetic (LAFM) state at $P_x$ = 0.5 $\sim$ 0.9 GPa~\cite{motoyama1,hassinger,amitsuka2,niklowitz,butch}. The bulk-SC state exists only below $P_x$\cite{amitsuka2,hassinger}.

 Previous studies have reported unusual electrical transport in URu$_2$Si$_2$~\cite{zhu,hassinger}. However, there is no consistency in the reported values of the exponent $n$ obtained from the fitting of the resistivity data with a general power law ${\,}{\rho_0}+AT^n$. This may come from a strong sample dependence of the electrical transport in URu$_2$Si$_2$ that  has been carefully studied in collaborative work between research groups in JAEA/Tokai and CEA/Grenoble~\cite{matsuda1,matsuda2}. The electrical resistivity of samples with different quality has been analyzed with the general power law just above $T_{sc}$ at ambient pressure. It was found that the value of the power law exponent $n$ depends on sample quality when the residual resistivity ratio RRR ( = ${\rho_{\rm RT}}/{\rho_{\rm 0}}$) (where ${\rho_{\rm 0}}$ and ${\rho_{\rm RT}}$ are a residual resistivity and the value of the resistivity at room temperature, respectively) is less than roughly 100. For RRRs above 100, the value of $n$ saturates to about 1.5 for the electrical resistivity with current along the $a$ and $c$-axes~\cite{matsuda2}.  We conclude that the deviation from the Fermi liquid theory is intrinsic to the electronic state of URu$_2$Si$_2$ at ambient pressure.  A further question is whether the ``non-Fermi liquid" behavior is intrinsic to the HO phase or not. In this letter, we report the detailed investigation on the electrical transport of URu$_2$Si$_2$ of an ultra-clean sample at high-pressure across $P_x$. We find a strong correlation between anomalous quasiparticle scattering and superconductivity in the HO phase.

    The high quality single crystal of URu$_2$Si$_2$ was grown by Czochralski pulling. The details of the sample preparation are given in the ref. 11. The electrical resistivity at high pressure was measured using the ac four terminal method in a $^3$He cryostat using a piston cylinder-type high-pressure cell. Daphne 7474 was used as pressure transmitting medium~\cite{murata,tateiwa}.  The viscosity of the Daphne 7474 is two orders of magnitude lower than that of the Daphne 7373 and a Fluorinert mixture~\cite{nakamura,varga}, commonly used in the previous studies~\cite{motoyama1,amitsuka2,niklowitz}. This indicates the better pressure-quality in the present study. It is difficult to estimate the RRR value because ${\rho_{\rm 0}}$ is negative if the resistivity just above $T_{sc}$ is simply extrapolated to 0 K. A lower value of the RRR was estimated as 300 using the resistivity value (${\rho_{T_{sc}}}$) just above $T_{sc}$ (RRR = ${\rho_{\rm RT}}/{\rho_{T_{sc}}}$). The residual resistivity is very small, far less than 1 ${\mu}{\Omega}{\cdot}$cm and the real RRR value exceeds 1000, indicating ultra-cleaness of the single crystal~\cite{matsuda1}.  We note that the quality of samples used in the previous studies is not enough to justify a conclusive discussion of the electrical transport in URu$_2$Si$_2$\cite{hassinger,butch,zhu}. The values of RRR are far less than 100. With the present sample with much higher RRR value, it becomes possible to investigate the intrinsic electrical transport under high pressure.
        
Figure 1 (a) shows the pressure-temperature phase diagram of URu$_2$Si$_2$ determined in the present resistivity measurement. Open squares indicate the transition temperature ${T_0}(P)$ or $T_{\rm N}(P)$ (at high pressure). Closed circles indicate the SC temperature ${T_{sc}}(P)$ below $P_x$ and open circles ${T_{sc}}(P)$ above the pressure that show the applied current dependence. 
 
 Fig. 1 (b) shows the low temperature electrical resistivity $\rho_a$ at 1 bar, 0.24, 0.54, and 0.75 GPa below $P_x$. At 1 bar, a clear SC transition was observed at $T_{sc}$ = 1.43 K. The transition temperature decreases with increasing pressure. The value of $T_{sc}$ is 0.94 K at 0.75 GPa. The resistivity value in the normal state decreases with increasing pressure, suggesting the suppression of the inelastic scattering of quasiparticles under compression.  Figure 1 (c) show the temperature dependences of $\rho_a$ at 0.94 and 1.51 GPa above $P_x$. The bulk-SC state exists only below $P_x$, while a broad SC transition in the resistivity is still observed in the pressure region far above $P_x$~\cite{amitsuka2,hassinger}. The persistence of the SC transition seen in the resistivity seems to be related to residual HO phase well beyond the LAFM boundary~\cite{amitsuka2,hassinger}. The filamentary SC above 0.94 GPa suggests that $P_x$ lies between 0.75 and 0.94 GPa in the present study. The shape of the phase boundary of $T{_x}(P)$ is sensitive to experimental conditions such as the pressure-transmitting medium\cite{niklowitz}. The line of $T{_x}(P)$ intersects the other two lines of $T{_0}(P)$ and $T{_{\rm N}}(P)$ at a tricritical point of $P^*$~\cite{hassinger}. In the ref. 6, $P^*$ was determined from the resistivity data around $T_0$ (or $T_{\rm N}$)~\cite{hassinger}. If we estimate it in the same way from the present data, $P^*$ is below 0.94 GPa and the phase boundary line of $T{_x}(P)$ runs upward with a very steep gradient as shown by the dotted line. Since the boundary is basically insensitive to the resistivity, it should be checked by a future study with other measurements such as a thermal expansion and heat capacity. 
\begin{figure}[t]
\begin{center}
\includegraphics[width=6.7cm]{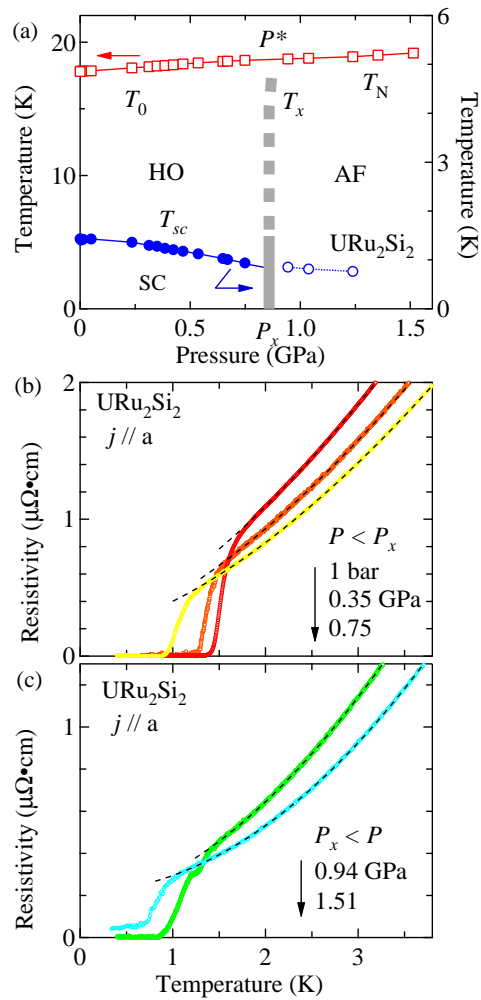}  
 \end{center}
\caption{\label{fig:epsart} (Color online) Pressure phase diagram determined in the present study of URu$_2$Si$_2$. Temperature dependences of the low temperature electrical resistivity $\rho_a$ at (b) 1 bar, 0.35, 0.75 GPa below $P_x$, and (c) 0.94, and 1.51 GPa above $P_x$ in URu$_2$Si$_2$. The dotted lines represent the fit of a generalized power law ${\rho}{\,}={\,}{\rho_0}+{\,}{A_n}T{^n}$ to the data from $T_l$ = $T_{sc}$(onset) +70 mK to 3.0 K where $T_{sc}$(onset) is the onset temperature of the SC transition in URu$_2$Si$_2$. }
\end{figure}

 \begin{figure}[t]
\begin{center}
\includegraphics[width=6.7cm]{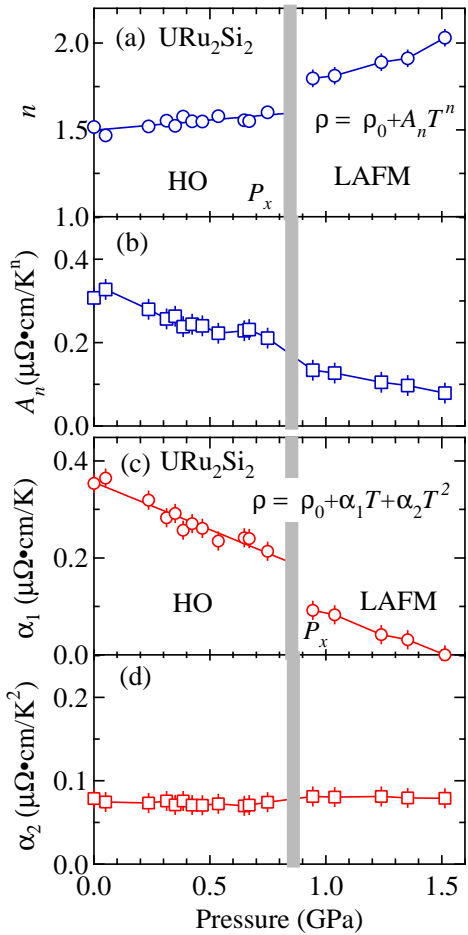}  
 \end{center}
\caption{\label{fig:epsart} (Color online) Pressure dependences of (a) the resistivity exponent $n$ and (b) the coefficient $A_n$ obtained by fitting the generalized power law ${\rho}{\,}={\,}{\rho_0}+{\,}{A_n}T{^n}$, and (c) the coefficient ${{\alpha}_1}$ and (d) ${{\alpha}_2}$ obtained by fitting the expression ${\rho}{\,}={\,}{\rho_0}+{\,}{{\alpha}_1}T+{\,}{{\alpha}_2}T{^2}$ to the resistivity in the temperature regions from $T_l$ to 3.0 K in URu$_2$Si$_2$.}
\end{figure}

 Firstly, the temperature dependence of ${\rho}_a$ at 1 bar was analyzed with the equation $\rho$ = ${\rho}_0$ + ${A_n}{T^n}$+$B$(1+2$T/{\Delta}$)exp(-${\Delta}/T$) in a wide temperature region from $T_l$ = $T_{sc}$(onset) +70 mK to 16 K, just below $T_0$. Here, $T_{sc}$(onset) is the onset temperature of the SC transition and ${\rho}_{sw}$ = $B$(1+2$T/{\Delta}$)exp(-${\Delta}/T$) is the contribution from an excitation with an gap $\Delta$ in the HO phase\cite{hassinger}. This contribution becomes negligibly small: ${{\rho}_{sw}}/{{\rho}}$ $<$ 10$^{-5}$ below 4 K. We ignore it in following analyses. 
 
 We analyzed the temperature dependence of $\rho_a$ with the generalized power law ${\rho}{\,}={\,}{\rho_0}+{\,}{A_n}T{^n}$ in the temperature region from $T_l$ to 3.0 K\cite{mathur}. The dotted lines in the figure are the results of the fit to the data. The pressure dependences of $n$ and $A_n$ obtained from the analyses in the temperature regions are shown in Figure 2 (a) and (b). Below $P_x$, $n$ shows a weak pressure dependence with approximately 1.5 $\pm$ 0.1. Above $P_x$, $n$ increases with increasing pressure and the value at 1.51 GPa becomes 2.0. There seems to be a discontinuous change in the pressure dependence of $n$ at $P_x$. The value of $A_n$ simply decreases with increasing pressure with no anomalous behavior in its pressure dependence. It is difficult to discuss the exponent $n$ in the LAFM state above $P_x$ since the filamentary superconductivity from the residual HO phase affects the analysis of the data in the pressure region close to $P_x$. The volume of the residual phase decreases with increasing pressure. Zero resistivity was not observed at 1.35 and 1.51 GPa.  Assuming that the LAFM phase is dominant at 1.51 GPa, the value of the power $n$ is 2.0, expected in Fermi liquid theory.

       \begin{figure}[t]
\begin{center}
\includegraphics[width=6.7cm]{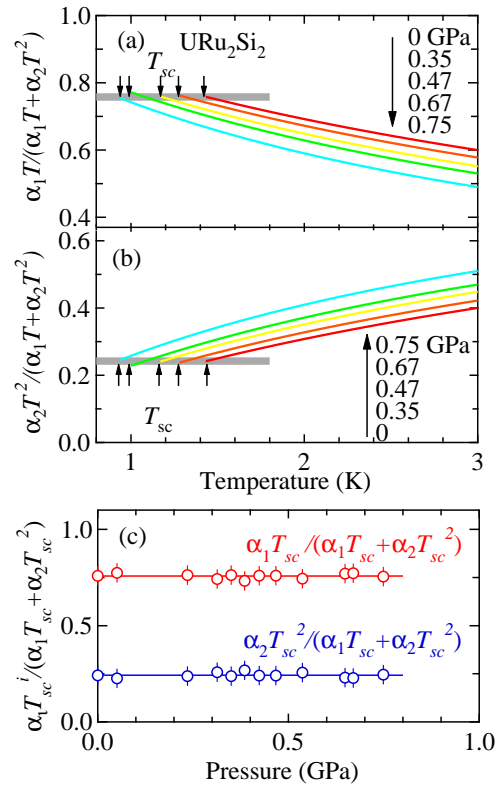}  
 \end{center}
\caption{\label{fig:epsart} (Color online) Temperature dependences of (a) ${{\alpha}_1}T/({{\alpha}_1}T+{{\alpha}_2}T{^2})$ and (b) ${{\alpha}_2}{T^2}/({{\alpha}_1}T+{{\alpha}_2}T{^2})$ calculated using the values of the coefficients ${{\alpha}_1}$ and ${{\alpha}_2}$ obtained from the fits of the data from $T_l$ to 3.0 K. (c) Pressure dependences of ${{\alpha}_1}{T_{sc}}/({{\alpha}_1}{T_{sc}}+{{\alpha}_2}{T_{sc}}{^2})$ and ${{\alpha}_2}{{T_{sc}}^2}/({{\alpha}_1}{T_{sc}}+{{\alpha}_2}{T_{sc}}{^2})$ in URu$_2$Si$_2$.}
\end{figure}

 The analysis with the generalized power law reveals the difference of the electrical transport between the HO and LAFM phases. On the contrary, recent Fermi surface studies under high pressure across $P_x$ suggest the similarity of the Fermi surface topology between the HO and LAFM phases~\cite{nakashima,hassinger2}. To gain further insights into the electrical transport in URu$_2$Si$_2$, we now analyze the data in the temperature regions from $T_l$ to 3.0 K using the expression ${\rho}{\,}={\,}{\rho_0}+{\,}{{\alpha}_1}T+{\,}{{\alpha}_2}T{^2}$ that has been used in the analysis of the anomalous electrical transport in the organic superconductors, the iron pnictide superconductors, and the high-$T_c$ cuprate superconductors~\cite{leyraud2,cooper,taillefer}. The expression assumes two independent scattering rates: one is an isotropic rate in $k$-space that gives the $T^2$ term in the resistivity and the other is anisotropic one that does the $T$-linear term.  Although the expression is empirical and further theoretical study is required for the justification of the application to URu$_2$Si$_2$, the following analysis gives interesting view points for the HO phase.

  The pressure dependencies of ${{\alpha}_1}$ and ${{\alpha}_2}$ are shown in Figure 2 (c) and (d). It is found that the contribution to the resistivity from the term ${{\alpha}_1}T$ is far larger than that from ${{\alpha}_2}T^{2}$ in the HO phase below $P_x$. With increasing pressure, the value of ${{\alpha}_1}$ decreases monotonously and shows a discontinuous decrease at $P_x$. Above $P_x$, ${{\alpha}_1}$ decreases strongly with increasing pressure going to 0 at 1.51 GPa. The value of ${{\alpha}_1}$ remains finite up to 1.35 GPa even though the ground state changes to the LAFM phase. This may be due to the residual HO phase in the LAFM phase as mentioned above. Compared with the drastic change of the ${{\alpha}_1}$, the coefficient ${{\alpha}_2}$ shows only weak pressure dependence, which is consistent with the fact that the Fermi surface topology does not change across the critical pressure $P_x$ studies~\cite{nakashima,hassinger2}. This suggests the validity of the analysis with the expression for the electrical transport in URu$_2$Si$_2$. We speculate that the scattering process of the quasiparticles on most regions of the Fermi surface obeys the Fermi liquid theory but the process in specific regions of the surfaces deviates from the theory in the HO phase.

\begin{figure}[t]
\begin{center}
\includegraphics[width=6.7cm]{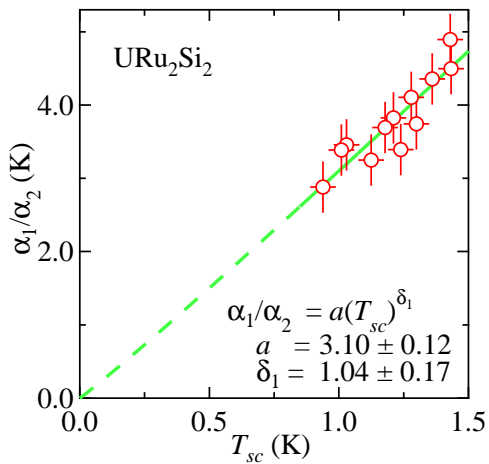}  
 \end{center}
\caption{\label{fig:epsart} (Color online) (a)Relation between the superconducting transition temperature $T_{sc}$ and ${{\alpha}_1}/{{\alpha}_2}$ in URu$_2$Si$_2$. The (dashed) line is a fit with the relation ${{\alpha}_1}/{{\alpha}_2}$ =  ${a}({T_{sc}})^{\delta_1}$, where the values of $a$ and ${\delta}_1$ are determined as 3.10 $\pm$ 0.12 and 1.04 $\pm$ 0.17, respectively,}
\end{figure}

  Figure 3 (a) and (b) show ${{\alpha}_1}T/({{\alpha}_1}T+{{\alpha}_2}T{^2})$ and ${{\alpha}_2}{T^2}/({{\alpha}_1}T+{{\alpha}_2}T{^2})$ that correspond to the ratios of the terms ${{\alpha}_1}T$ and ${{\alpha}_2}{T^2}$ to the resistivity due to the electron correlations ${\Delta}{\rho}$ (= ${\rho}$ - ${\rho}_0$). The contributions from the two terms to ${\Delta}{\rho}$ are comparable around 3 K. The contribution from ${{\alpha}_1}T$ to ${\Delta}{\rho}$ increases with decreasing pressure and becomes dominant just above $T_{sc}$. Interestingly, ${{\alpha}_1}T/({{\alpha}_1}T+{{\alpha}_2}T{^2})$ and ${{\alpha}_2}{T^2}/({{\alpha}_1}T+{{\alpha}_2}T{^2})$ show almost pressure-independent values of 0.74 $\pm$ 0.05 and 0.25 $\pm$ 0.05, respectively, just above $T_{sc}$ as shown in Fig. 3 (c). This suggests the relation ${{{\alpha}_1}/{\alpha_2}}{\,}{\propto}{\,}{T_{sc}}$. 
   
    Figure 4 shows the relation between $T_{sc}$ and ${{\alpha}_1}/{{\alpha}_2}$. The (dashed) line is a fit with the relation ${{\alpha}_1}/{{\alpha}_2}$ =  ${a}({T_{sc}})^{\delta_1}$, where the values of $a$ and ${\delta}_1$ are determined as 3.10 $\pm$ 0.12 and 1.04 $\pm$ 0.17, respectively, suggesting a linearity between $T_{sc}$ and ${{\alpha}_1}/{{\alpha}_2}$. Since the pressure dependence of the coefficient ${{\alpha}_2}$ is very weak as shown in Fig. 3 (c), the value of $T_{sc}$ depends primarily on the coefficient ${{\alpha}_1}$. We obtain the relation ${{\alpha}_1}$ =  ${c}{T_{sc}}^{{\delta}_2}$, where $c$ and ${\delta}_2$ are determined to be 0.22 $\pm$ 0.01 and 1.11 $\pm$ 0.15, respectively, suggesting an almost linear relation between ${{\alpha}_1}$ and ${T_{sc}}$. These results suggest the strong correlation between anomalous quasiparticle scattering and unconventional superconductivity in the HO phase of URu$_2$Si$_2$. We suggest that the anomalous quasiparticles scattering and the superconductivity have a common origin, possibly rooted in the magnetic excitations at $Q_{0}$ since the excitations are observed only in the HO phase and peak position $E_0$ in the inelastic resonance is shifted below $T_{sc}$~\cite{villaume,bourdarot2}. It is noted that the almost same results are obtained when the the resistivity data between $T_l$ to 3.4 K are analyzed.

   Similar correlation between the $T$-linear resistivity and $T_{sc}$ has been found in the organic superconductors, the iron pnictide superconductors and the high-$T_c$ cuprate superconductors as summarized by Taillefer~\cite{taillefer}. This correlation may be a universality in the unconventional superconductors. In these systems, the $T$-linear resistivity appearing around a magnetic phase boundary has been interpreted as manifestation of quantum criticality. We suggest that the HO phase in URu$_2$Si$_2$ shares this universality inherent to the strong correlated electron superconductors near the quantum criticality, although the peculiarity of the HO phase originating from the multipole degree of freedom in the 5 $f$ electrons has been stressed in theoretical studies without an experimental evidence\cite{mydosh}. The ``hidden order" could exist in other strongly correlated electron compounds. Unidentified mysterious phases have been reported near the SC phase or a quantum critical point such as ``pseudo-gap phase" in the cuprate superconductors~\cite{taillefer}. It is interesting to note recent studies on the unconventional superconductors including URu$_2$Si$_2$ from the view point of electronic nematicity whose emergence around a quantum critical point has been studied theoretically\cite{okazaki,chu,lawler,sachdev}. We hope this study provides a different point of view for existing or future theories of the HO phase.

   In summary, our analysis of resistivity measurements near $T_{sc}$ on very high resistance ratio URu$_2$Si$_2$ single crystal now places this unusual material in the group of cuprate, pnictide and organic quantum critical superconductors identified by Tailleferfs resistivity correlation. We find the ratio of the coefficient of the linear in $T$ term  in the resistivity to the quadratic term $T^2$ ${{\alpha}_1}/{{\alpha}_2}$ is directly proportional to $T_{sc}$. Since the value of ${\alpha}_2$ is almost pressure-independent, ${{\alpha}_1}$ is roughly proportional to $T_{sc}$. Similar correlation has been reported in other strongly correlated electron superconductors. 
   
    We thank Kamran Behnia (ESPCI, Pari) for enlightening suggestions. This work was supported by a Grant-in-Aid for Scientific Research on Innovative Areas ``Heavy Electrons (No. 20102002, No. 23102726), Scientific Research S (No. 20224015), A(No. 23246174), C (No. 22540378), Specially Promoted Research (No. 20001004) and Osaka University Global COE program (G10) from the Ministry of Education, Culture, Sports, Science and Technology (MEXT) and Japan Society of the Promotion of Science (JSPS).

\bibliography{apssamp}

\begin{thebibliography}{99} 


\bibitem{palstra}T. T. M. Palstra, A. A. Monovsky, J. van den Berg, A. J. Dirkmaat, P. H. Kes, G. J. Nieuwenhuys, and J. A. Mydosh Phys. Rev. Lett. {\bf 55}, 2727 (1985).
\bibitem{kasahara}Y. Kasahara, T. Iwasawa, H. Shishido, T. Shibauchi, K. Behnia, Y. Haga, T. D. Matsuda, Y.  {\=Onuki}, M. Sigrist, and Y. Matsuda, Phys. Rev. Lett. {\bf 99}, 116402 (2007). 


\bibitem{mydosh}For a review, see J. A. Mydosh and P. Oppeneer, Rev. Mod. Phys. {\bf 83}, 1301 (2011).

\bibitem{motoyama1}G. Motoyama, T. Nishioka, and N. K. Sato,  Phys. Rev. Lett. {\bf 90}, 166402 (2003).

\bibitem{amitsuka2} H. Amitsuka, K. Matsuda, I. Kawasaki, K. Tenya, M. Yokoyama, C. Sekine, N. Tateiwa, T. C. Kobayashi, S. Kawarazaki, and H. Yoshizawa, J. Magn. Magn. Mater. {\bf 310}, 214 (2007).
\bibitem{hassinger}E. Hassinger, G. Knebel, K. Izawa, P. Lejay, B. Salce, and J. Flouquet, Phys. Rev. B {\bf 77}, 115117 (2008).
\bibitem{niklowitz}P. G. Niklowitz, C. Pfleiderer, T. Keller, M. Vojta, Y.-K. Huang, and J. A. Mydosh, Phys. Rev. Lett. {\bf 104}, 106406 (2010).
\bibitem{butch}N. P. Butch, J. R. Jeffries, S. Chi, J. B. Le{\~ a}o, J. W. Lynn, and M. B. Maple, Phys. Rev. B {\bf 82}, 060408 (2010).
\bibitem{zhu}Z. Zhu, E. Hassinger, Z. Xu, D. Aoki, J. Flouquet, and K. Behnia, Phys. Rev. B {\bf 80}, 172501 (2009).


\bibitem{matsuda1}T. D. Matsuda, D. Aoki, S. Ikeda, E. Yamamoto, Y. Haga, H. Ohkuni, R. Settai and Y. {\=Onuki}: J. Phys. Soc. Jpn. {\bf 77} 362 (2008).
\bibitem{matsuda2}T. D. Matsuda, E. Hassinger, D. Aoki, V. Taufour, G. Knebel, N. Tateiwa, E. Yamamoto, Y. Haga, Y. {\=Onuki}, Z. Fisk, and J. Flouquet, J. Phys. Soc. Jpn. {\bf 80} 114710 (2011). 
\bibitem{murata}K. Murata, K. Yokogawa, H. Yoshino, S. Klotz, P. Munsch, A. Irizawa, M. Nishiyama, K. Iizuka, T. Nanba, T. Okada, Y. Shirage, and S. Aoyama,  Rev. Sci. Instrum. {\bf 79}, 085101 (2008).
\bibitem{tateiwa}N. Tateiwa and Y. Haga, Rev. Sci. Instrum. {\bf 80}, 123901 (2009).
\bibitem{nakamura}Y. Nakamura, A. Takimoto and M. Matsui, J Phys: Conf. Ser. {\bf 215}, 012176 (2010).
\bibitem{varga}T. Varga, A. P. Wilkinson, and R. J. Angel,  Rev. Sci. Instrum. {\bf 74}, 4564 (2003).

\bibitem{mathur} N. D. Mathur, F. M. Grosche, S. R. Julian, I. R. Walker, D. M. Freye, R. K. W. Haselwimmer, and G. G. Lonzarich1, Nature {\bf 394}, 39 (1998).


\bibitem{nakashima}M. Nakashima, H. Ohkuni, Y. Inada, R. Settai, Y. Haga, E. Yamamoto, and Y. {\=Onuki}, J. Phys.: Condens. Matter {\bf 15}, S2011 (2003).
\bibitem{hassinger2}E. Hassinger, G. Knebel, T. D. Matsuda, D. Aoki, V. Taufour, and J. Flouquet, Phys. Rev. Lett. {\bf 105}, 216409 (2010).

\bibitem{leyraud2}N. Doiron-Leyraud, P. Auban-Senzier, S. R. de Cotret, C. Bourbonnais, D. J{\'e}rome, K. Bechgaard, and L. Taillefer,  Phys. Rev. B {\bf 80}, 214531 (2010).
\bibitem{cooper} R. A. Cooper, Y. Wang, B. Vignolle, O. J. Lipscombe, S. M. Hayden, Y. Tanabe, T. Adachi, Y. Koike, M. Nohara, H. Takagi, C. Prooust, and N. E. Hussey, Science {\bf 323}, 603 (2009).
\bibitem{taillefer} L. Taillefer, Annual Review of Condensed Matter Physics {\bf 1}, 51 (2010).



\bibitem{villaume}A. Villaume, F. Bourdarot, E. Hassinger, S. Raymond, V. Taufour, D. Aoki, and J. Flouquet, Phys. Rev. B {\bf 78}, 012504 (2008).
\bibitem{bourdarot2}F. Bourdarot, E. Hassinger, S. Raymond, D. Aoki, V. Taufour,  and J. Flouquet, J. Phys. Soc. Jpn. {\bf 79}, 094706 (2010).




\bibitem{okazaki}R. Okazaki, T. Shibauchi, H. J. Shi, Y. Haga, T. D. Matsuda, E. Yamamoto, Y. {\=Onuki}, H. Ikeda, and Y. Matsuda, Science {\bf 331}, 439 (2011).
\bibitem{chu}J-H. Chu, J. G. Analytis, K. D. Greve, P. L. McMahon, Z. Islam, Y. Yamamoto, and I. R. Fisher, Science {\bf 329}, 824 (2010).

\bibitem{lawler}M. J. Lawler {\it et al.}, Nature {\bf 466}, 347 (2010).
\bibitem{sachdev}E. G. Moon and S. Sachdev, Phys. Rev. B {\bf 82}, 104516 (2010).



\end{thebibliography}

\end{document}